%% file: main.tex
\documentclass[%
 reprint,
 amsmath,amssymb,
 aps,
]{revtex4-1}    

\usepackage{graphicx}
\usepackage{dcolumn}
\usepackage{bm}
\usepackage{float}
\usepackage{physics}
\usepackage{tikz}
\usetikzlibrary{calc}
\usepackage{multirow}
\usepackage{xcolor}
\usepackage[toc,page]{appendix}

\begin{document}


\title{Spin splitting and spin Hall conductivity in buckled monolayers of the group 14:\\ First-principles calculations}

\author{S. M. Farzaneh}
\email{farzaneh@nyu.edu}
\affiliation{Department of Electrical and Computer Engineering, New York University, Brooklyn, New York 11201, USA}
\author{Shaloo Rakheja}
\email{rakheja@illinois.edu}
\affiliation{Holonyak Micro and Nanotechnology Laboratory University of Illinois at Urbana-Champaign, Urbana, Illinois 61801, USA}

\begin{abstract}
Elemental monolayers of the group 14 with a buckled honeycomb structure, namely silicene, germanene, stanene, and plumbene, are known to demonstrate a spin splitting as a result of an electric field parallel to their high symmetry axis which is capable of tuning their topological phase between a quantum spin Hall insulator and an ordinary band insulator. 
We perform first-principles calculations based on the density functional theory to quantify the spin-dependent band gaps and the spin splitting as a function of the applied electric field and extract the main coefficients of the invariant Hamiltonian. 
Using the linear response theory and the Wannier interpolation method, we calculate the spin Hall conductivity in the monolayers and study its sensitivity to an external electric field.
Our results show that the spin Hall conductivity is not quantized and in the case of silicene, germanene, and stanene degrades significantly as the electric field inverts the band gap and brings the monolayer into the trivial phase. 
The electric field induced band gap does not close in the case of plumbene with a spin Hall conductivity that is robust to the external electric field. 
\end{abstract}

\maketitle

\section{Introduction}
The buckled monolayers of the group 14 possess a topological phase known as the quantum spin Hall insulator \cite{kane_quantum_2005}.
An external electric field, via a substrate or a gate contact,  provides an experimentally feasible way to tune the spin properties of these materials which is highly desirable in spintronics applications \cite{tsai_gated_2013}. 
Calculations show that an external electric field is capable of switching the topological phase into a trivial band insulating phase \cite{ezawa_monolayer_2015}. Moreover, it is well known \cite{geissler_group_2013, ezawa_monolayer_2015} that the electric field induces spin splitting in the bands at the $K$ point of the Brillouin zone. 
However, the impact of the electric-field-induced spin splitting on the quantum spin Hall phase and the spin Hall conductivity is relatively less explored. 
With the recent experimental realization of stanene \cite{zhu_epitaxial_2015} and plumbene \cite{yuhara_graphenes_2019}, it is desirable to investigate the impact of the electric field on the spin splitting and consequently the spin Hall conductivity via systematic first-principles calculations based on the density functional theory and the linear response theory. 

Elemental monolayers of the group 14 of the periodic table are two-dimensional (2D) crystals of carbon, silicon, germanium, tin, and lead, which are known as graphene, silicene, germanene, stanene, and plumbene, respectively.
This family of 2D materials possess a variety of properties ranging from Dirac energy dispersion of the low energy excitations at the $K$ point, spin-orbit induced band gap,  possibility of quantum spin Hall insulating phase which is tunable via an external electric field~\cite{molle_buckled_2017}. 
Following the successful isolation of graphene a decade ago~\cite{novoselov_two-dimensional_2005}, other members of this family came into existence starting from silicene~\cite{lalmi_epitaxial_2010, vogt_silicene_2012, fleurence_experimental_2012, lin_structure_2012}, which was followed by germanene~\cite{davila_germanene_2014, lin_single-layer_2018} and more recently by stanene~\cite{zhu_epitaxial_2015, deng_epitaxial_2018, yuhara_large_2018} and plumbene~\cite{yuhara_graphenes_2019}.
In their most stable configuration, they show a buckled honeycomb structure which possesses a lower symmetry than that of graphene. 
Therefore, in general, the electronic states are expected to have lower number of degeneracies. 
Since, theses monolayers consist of heavier elements than carbon, spin-orbit coupling is expected to affect the electronic band structure more significantly. 
In fact, as one goes from silicene to plumbene, the intrinsic band gap at the $K$ point increases and the Fermi velocity decreases. 

Spin-orbit coupling in the buckled monolayers is of essential importance both in fundamental physics such as the quantum anomalous Hall effect \cite{wu_prediction_2014, yu_electric_2015} and the quantum spin Hall effects \cite{xu_large-gap_2013, ezawa_monolayer_2015} and also in spintronics applications such as the spin-polarized transistor \cite{pournaghavi_extrinsic_2017}, spin-valley logic \cite{ezawa_spin_2013, tao_two-dimensional_2019, tao_valley-spin_2020}, spin filter~\cite{tsai_gated_2013, rachel_giant_2014}, valley-polarized metal \cite{ezawa_valley-polarized_2012}, or the electrical switching of magnetization via spin-orbit torques \cite{li_valley-dependent_2016}. 
There exists rich literature on theoretical works on buckled monolayers in the past decade based on group theoretic works~\cite{geissler_group_2013, gert_effective_2016, kochan_model_2017},
effective Hamiltonians \cite{liu_low-energy_2011, liu_quantum_2011, ezawa_monolayer_2015, lew_yan_voon_effective_2015,  kochan_model_2017, li_spin_2018}, and first-principles calculations based on the density functional theory \cite{takeda_theoretical_1994,  cahangirov_two-_2009, sahin_monolayer_2009, liu_low-energy_2011, liu_quantum_2011,  drummond_electrically_2012, ni_tunable_2012, matthes_massive_2013, tsai_gated_2013, rivero_stability_2014, rahman_distortion_2014, zhao_first-principles_2016, yu_normal_2017, yang_can_2018, kurpas_spin-orbit_2019, mahmood_structural_2020}. 
Nevertheless, only few first-principles works \cite{drummond_electrically_2012, ni_tunable_2012, tsai_gated_2013, rahman_distortion_2014} study the effect of the electric field in buckled monolayers. Of these studies only Refs. \cite{drummond_electrically_2012, tsai_gated_2013} include the relativistic spin-orbit coupling effects. 
Ab initio studies \cite{matthes_intrinsic_2016,  matusalem_quantization_2019, matthes_influence_2014} on the spin Hall conductivity of germanene and stanene have only recently become available. 
Here, we go a step further by analyzing the electronic properties of all the monolayers: silicene, germanene, stanene, and plumbene. For each monolayer, we 
systematically investigate the effect of electric-field-induced spin splitting in conjunction with the spin Hall conductivity via fully relativistic first-principles calculations for a wider range of electric fields and energies than previous works \cite{tsai_gated_2013, matusalem_quantization_2019}. 
We quantify various spin properties of the buckled monolayers such as the spin splitting, spin-dependent band gaps, critical electric field required for topological phase transition, coefficients of the low energy invariant Hamiltonian, and the spin Hall conductivity with and without the electric field. 

A brief group theoretic analysis of the bands at the $K$ point is provided in Sec. \ref{sec:symm} where we show how the symmetry classification of the bands changes as the spin-orbit coupling, the buckling, and the electric field are introduced one by one. 
First-principles calculations of the band structure and the spin-split bands of the monolayers are studied in detail in Sec. \ref{sec:abinitio} where the effect of the external electric field is taken into account in the self-consistent  Kohn-Sham equations. 
The coefficients of an invariant Hamiltonian describing the low energy band structure at the $K$ point are extracted in Section \ref{sec:invariant}. 
The spin Hall conductivity of the monolayers is calculated and discussed in Sec. \ref{sec:shc}.
The paper concludes with a summary of key findings and outlook in Sec.~\ref{sec:conclusions}.

\section{\label{sec:symm}Crystal structure and symmetries}
\begin{figure}[ht]
  \centering
  \includegraphics[]{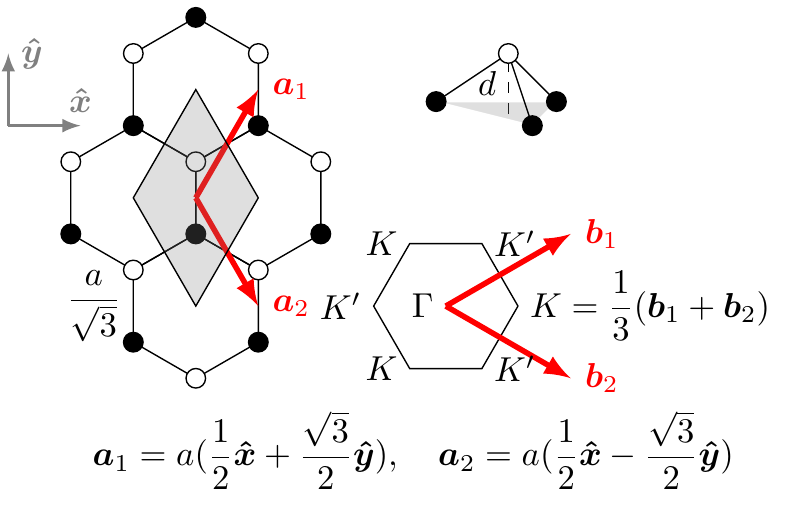}
  \caption{A two-dimensional buckled honeycomb crystal with the lattice constant $a$ and the buckling size $d$. The primitive unit cell is shaded in gray and the primitive vectors are drawn in red. Dark and light circles denote atoms on different sublattices. The first Brillouin zone along with the reciprocal vectors are shown on the bottom right.}
  \label{fig:crystal}
\end{figure}

The crystal structure of the monolayers of the group 14 consist of atoms arranged in a honeycomb structure as is shown in Fig. \ref{fig:crystal}. 
The honeycomb structure can be thought of as a Bravais lattice with a basis of two atoms, i.e., dark and light circles in the figure. 
The dark and light circles each consist a triangular Bravais lattice by themselves. 
In general, there can exist an out-of-plane buckling in the honeycomb structure depending on the relative stability of $sp^2$ and $sp^3$ hybridizations as well as the exchange-correlation potentials \cite{takeda_theoretical_1994}.  
The buckling size, denoted by $d$, the distance between the two triangular sublattices. 
The buckling is zero in the case of planar graphene but increases monotonically as one goes from the lighter to the heavier elements of the group 14. 
The first Brillouin zone of the honeycomb structure is also shown in Fig.~\ref{fig:crystal} where the nonequivalent high symmetry points on the boundary of the Brillouin zone, $K$ and $K'$, host the low energy excitations. 

\begin{figure}[ht]
  \centering
  \includegraphics[]{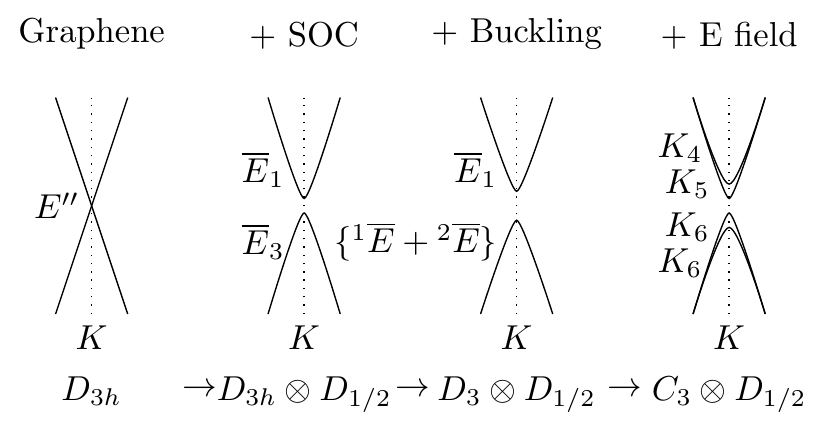}
  \caption{Symmetry classification of the Dirac point with respect to the group of the wave vector at the $K$ point. As one goes from spinless graphene to the buckled monolayers in the presence of the spin-orbit coupling and the external electric field, different symmetries are introduced or broken and consequently degeneracies and symmetry properties change.}
  \label{fig:symm}
\end{figure}
The symmetry of the planar graphene is classified by the symmorphic space group $P6/mmm$ (\#191) which is homomorphic to the point group $D_{6h}$. 
The buckled monolayers have a lower symmetry than that of graphene because the 6-fold rotational symmetry reduces to a 3-fold one and the horizontal mirror symmetry is broken as well. However, they retain the inversion symmetry. 
Their symmetry is classified by the space group $P\overline{3}m1$ (\#164) with the corresponding point group $D_{3d}$ which is a subgroup of $D_{6h}$. 
Although Ref. \cite{kochan_model_2017} studies the global symmetry properties of graphene systems, here we focus on the symmetry properties of the $K$ point in the Brillouin zone. 

Group theoretic arguments can provide insight into the qualitative behavior of the bands such as the number of degeneracies and how or if they are lifted as the symmetry is lowered. 
The details of the symmetry analysis and the symmetry tables are provided as the supplemental material \footnote{See Supplemental Material at [URL will be inserted by publisher] for the details of the symmetry analysis and the symmetry tables which includes Reference \cite{dresselhaus_group_2008}.}. 
Here the symmetry properties are briefly mentioned. 
Figure \ref{fig:symm} illustrates the qualitative behaviour of the bands in the vicinity of the $K$ point. 
It illustrates how the irreducible representations (IR) of the bands at the Dirac point change. 
Specifically, 
the 
degeneracies are lifted as one goes from graphene to other lower symmetry materials in the presence of buckling and an external electric field.
Each step is denoted by a point group corresponding to the group of the wave vector at the $K$ point.
Starting from the spinless graphene with the point group $D_{3h}$ the Driac point is labeled with $E''$, a two-dimensional IR which implies a 2-fold degeneracy at the Dirac point \cite{kogan_symmetry_2013}. 
The effect of the spin-orbit coupling can be shown by using the double group representations which are obtained by the direct product of the point group with the spinor representation $D_{1/2}$, i.e., $D_{3h}\otimes D_{1/2}$. 
The IRs of this double group are at most two-dimensional. Therefore, no 4-fold degeneracy (including spins) is allowed at the Dirac point and therefore a gap opens up. 
After the inclusion of spinors with the representation $\overline{E}$, the IR of the bands at the Dirac point changes to $E''\otimes \overline{E} = \overline{E}_1 + \overline{E}_3$ which are two 2-fold bands. 
The buckling breaks some of the symmetries of the planar graphene structure such as the in-plane mirror symmetry and the 6-fold rotational symmetries resulting in point group $D_3$. No degeneracy is lifted as the buckling is introduced and the bands are represented with the same 2-fold degeneracy but with labels according to point group $D_3$, i.e., $\overline{E}_1 (D_{3h})\rightarrow \overline{E}_1 (D_3)$ and $\overline{E}_3 (D_{3h})\rightarrow \{{}^1\overline{E}+{}^2\overline{E}\}(D_3)$. 
The electric field reduces the symmetry further to the point group $C_3$ which contains only one-dimensional representations and therefore all degeneracies are lifted. 
Although group theory can predict the spin splitting, the ordering and the energy of the bands are only obtained through calculations or experiments. 

\section{\label{sec:abinitio}First-principles calculations}
First-principles calculations based on the density functional theory are performed via the Quantum ESPRESSO  suite \cite{giannozzi_quantum_2009, giannozzi_advanced_2017}. 
The projector augmented wave method \cite{blochl_projector_1994} which generalizes the pseudopotential method is used to improve the computational efficiency. The pseudopotential files are obtained from Ref. \cite{dal_corso_pseudopotentials_2014}. 
Scalar-relativistic pseudopotentials are used for structural relaxation whereas fully-relativistic pseudopotentials are used to capture spin-orbit coupling effects. 
The exchange-correlations functional utilizes the generalized gradient approximation \cite{perdew_generalized_1996}. 
Other parameters and details of the first-principles setup are presented in Table \ref{tb:abinitio-setup}. 
\bgroup
\def\arraystretch{1.5}
\begin{table}[ht]
\centering
\caption{Parameters in the setup of the first-principles calculations.}
\label{tb:abinitio-setup}
\begin{tabular}{p{1.8in}ccccc}
\hline
& C & Si & Ge & Sn & Pb \\
\hline
Wavefunction $E_\text{cut}$ (Ry)& 40 & 45 & 50 & 70 & 46\\
Charge density $E_\text{cut}$ (Ry)& 326 & 180 & 200 & 280 & 211\\
Number of bands & 16 & 16 & 16 & 36 & 36\\
Energy convergence threshold & \multicolumn{5}{c}{$10^{-8}$   a.u.}\\
Force convergence threshold & \multicolumn{5}{c}{$10^{-7}$  a.u.}\\
Number of Wannier orbitals & \multicolumn{5}{c}{$8$}\\
\hline
Calculation & \multicolumn{5}{c}{\# of $k$ points} \\
\hline
Structural optimization & \multicolumn{5}{c}{$12\times12\times1$} \\
Self consistent field & \multicolumn{5}{c}{$48\times48\times1$} \\
Band structure $k$ path & \multicolumn{5}{c}{150} \\
Density of states & \multicolumn{5}{c}{$96\times96\times1$} \\
Wannier interpolation & \multicolumn{5}{c}{$800\times800\times1$} \\
\hline
\end{tabular} 
\end{table}

The lattice parameters are obtained by the structural optimization using the Broyden–Fletcher–Goldfarb–Shanno algorithm which is a quasi-Newton optimizer where the forces are calculated using the Hellmann-Feynman theorem. 
The optimized lattice parameters, such as the lattice constant and the buckling height, which minimize the total force and stress are listed in Table \ref{tb:structural-params}. 
\def\arraystretch{1.5}
\begin{table}[b]
\centering
\caption{Optimized structural parameters of graphene, silicene, germanene, stanene, and plumbene, i.e., the lattice constant $a$ and the buckling size $d$ as defined in Fig. \ref{fig:crystal}.}
\label{tb:structural-params}
\begin{tabular}{cccccc}
\hline
& C & Si & Ge & Sn & Pb \\
\hline
a (\AA) & 2.466 & 3.868 & 4.022 & 4.652 & 4.924 \\
d (\AA) & 0.000 & 0.453 & 0.687 & 0.862 & 0.939 \\
\hline
\end{tabular} 
\end{table}
As seen from the table, the values are similar to the ones reported previously in the literature \cite{cahangirov_two-_2009, liu_low-energy_2011, tsai_gated_2013}.
The supercell contains a 20 \AA~ vacuum to avoid fictitious interlayer interaction due to the periodic boundary condition. 
The external electric field is modeled as an effective saw-like potential which is added directly to the self consistent Kohn-Sham equations. 
The magnitude of the homogeneous electric field, which is perpendicular to the crystal plane, is increased gradually to ease the convergence. 
The pressure of the crystal is kept below 0.5 kbar. The non-zero pressure is mostly due to the periodicity in the direction perpendicular to the crystal plane and can be reduced by increasing the vacuum size further. However, its effect on the optimized lattice constant is negligible, i.e. 0.001 \AA~ for a vacuum twice as large. 

\begin{figure}
    \includegraphics[]{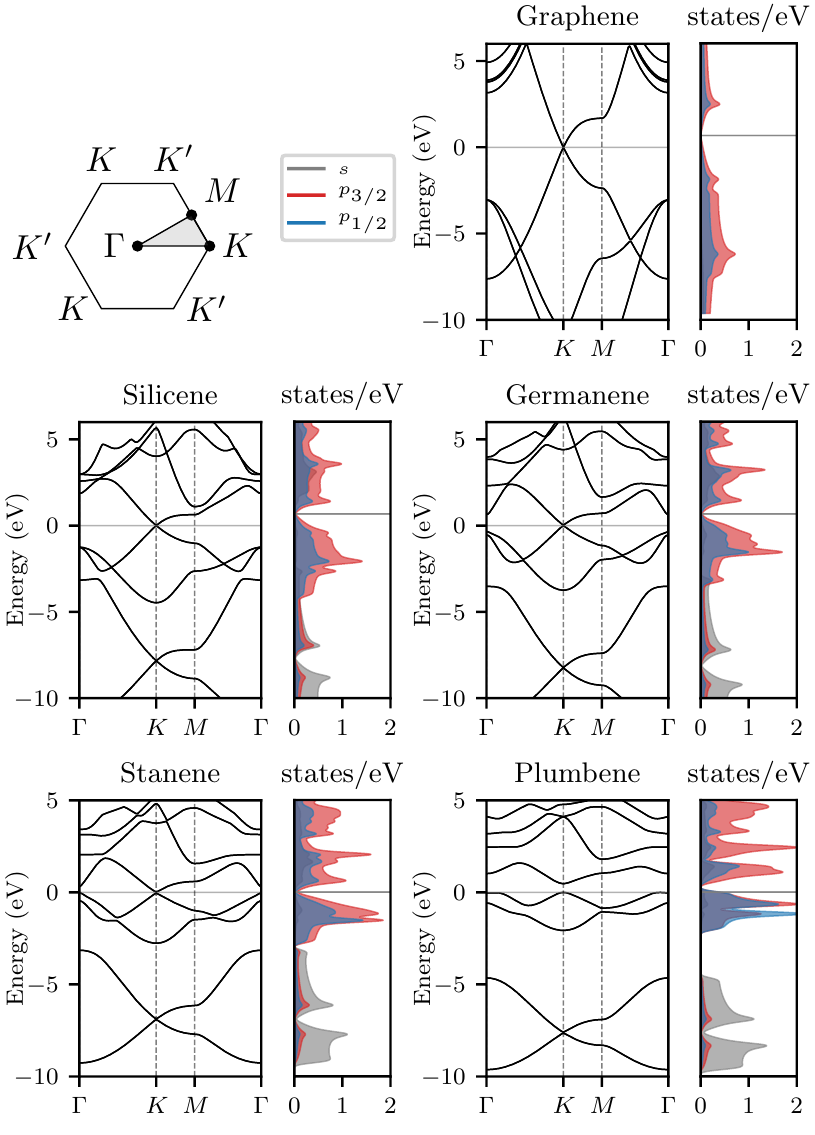}
    \caption[example] 
    {The band structure of graphene, silicene, germanene, stanene, and plumbene along with their corresponding density of states (DOS) in 1/eV units. The energy axis is relative to the Fermi energy denoted by the gray horizontal line.}
    \label{fig:band} 
\end{figure}  

The fully-relativistic band structure of graphene, silicene, germanene, and stanene along with their density of states are illustrated in Fig. \ref{fig:band}. 
The bands shown in the energy window are composed of only $s$ and $p$ orbitals as the $d$ orbitals, in the case of germanene, stanene, and plumbene, are narrow and localized far below the Fermi energy. 
The density of states approaches zero at the Fermi energy due to the band gap caused by the spin-orbit coupling effects. 
The density of states of graphene is more dispersed due to its higher bandwidth compared to other monolayers. 
The band gap of graphene is very small, of the order of $\mu$eV, due to the weak atomic spin-orbit coupling of carbon atoms. 
The resulting band gap of silicene, germanene, stanene, and plumbene are 1.5 meV, 23.7 meV, 77.2 meV, and 477 meV, respectively. 
There is a good match between these values and the ones reported before \cite{liu_low-energy_2011, tsai_gated_2013}. 

\begin{figure}
    \includegraphics[]{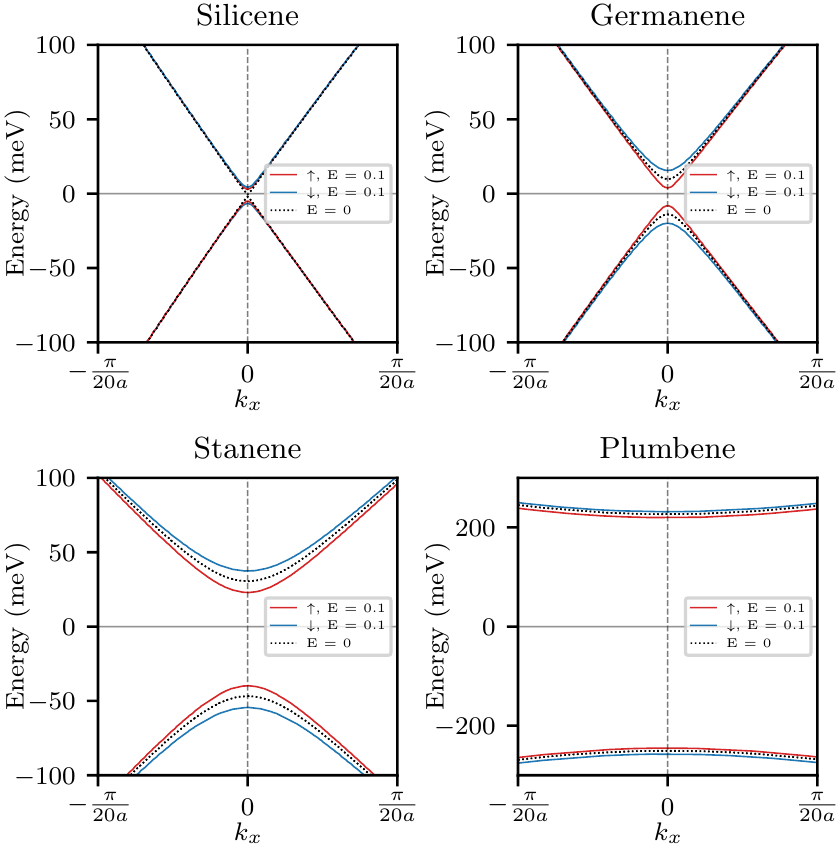}
    \caption[example] 
    {The effect of the electric field on the band structure of silicene, germanene, stanene, and plumbene at the $K$ point of the Brillouin zone along the $x$ direction. The energy axis is relative to the Fermi energy denoted by the gray horizontal line. 
    The spin splitting and the change in the band gap depend on the relative strength of the electric field and the spin-orbit coupling. The electric field is in V/\AA~ units.}
    \label{fig:band-efield} 
\end{figure}  
In the presence of an electric field the inversion symmetry of the crystal is broken and the degeneracies of the bands at the $K$ and the $K'$ points are lifter. Figure \ref{fig:band-efield} depicts the spin-split bands in the vicinity of the $K$ point with and without an external electric field. 
The energy window for plumbene is chosen larger than the rest to capture the band gap.
The band gap without an electric field is a result of the intrinsic spin-orbit coupling which is stronger for the monolayers with heavier elements. 
We note that the energy band diagram looks the same for the $K'$ point except that the spins are opposite to that at the $K$ point due to the time reversal symmetry, i.e., $E_\uparrow(K)=E_\downarrow(K')$. 

\begin{figure}
    \includegraphics[]{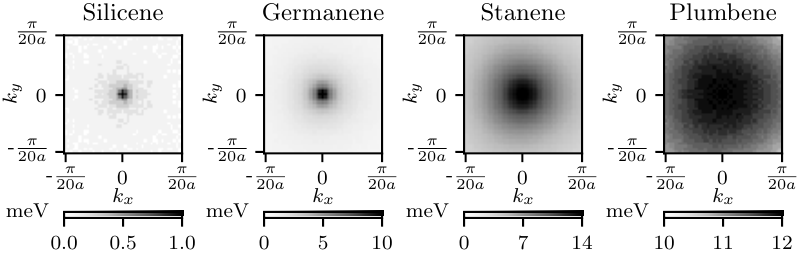}
    \caption[example] 
    {The spread of the spin splitting induced by the electric field ($E$=0.1 V/\AA) in the vicinity of the $K$ point for silicene, germanene, stanene, and plumbene.}
    \label{fig:spread} 
\end{figure}  
The spin splitting is not uniform across the $k$ space. It peaks at the $K$ point and decays outward. The spread of the splitting is also not the same for different monolayers. Figure \ref{fig:spread} shows the spread of spin splitting close to the $K$ point in the presence of an electric field with a magnitude of $0.1$ V/\AA. As seen from the figure the heavier the element, the stronger the spin-orbit coupling and the larger the band gap is which in turn corresponds to a heavier effective mass and therefore a more spread spin splitting. 

\begin{figure}
     \includegraphics[]{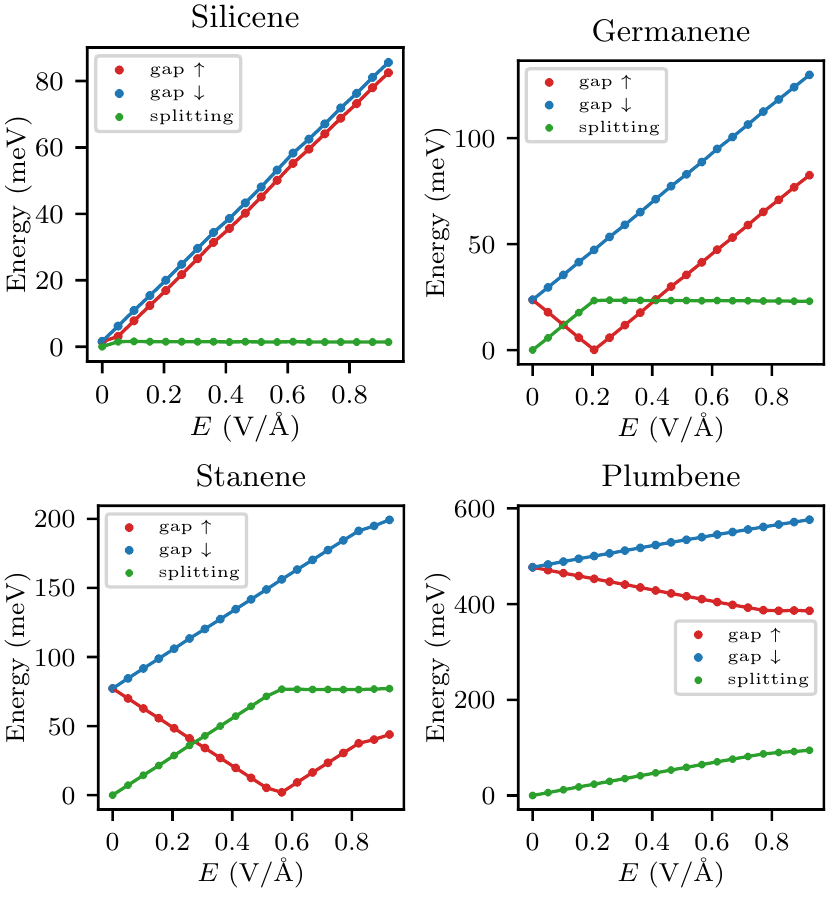}
     \caption[example] 
    {The spin-dependent band gap and the spin splitting as a function of the external electric field for silicene, germanene, stanene, and plumbene. The spin-up band gap closes at the critical electric field. Germanene and stanene show a band inversion at the critical field whereas no transition happen for silicene and plumbene.}
    \label{fig:splitting} 
\end{figure}  
To investigate the impact of the electric field more closely we calculate the band structure for a wide range of electric field magnitudes. 
Figure \ref{fig:splitting} plots the spin-up and spin-down band gaps along with the spin splitting of each band as a function of the electric field. 
The general behavior of the spin-dependent band gap is that the spin-down band gap linearly increases whereas the spin-up band gap initially decreases until it reaches zero, or the valley polarized metal state \cite{ezawa_valley-polarized_2012}, and then increases again. 
It has been shown \cite{ezawa_monolayer_2015} that a topological phase transition occurs at the zero gap point therefore changing the topological phase from a quantum spin Hall insulator to an ordinary band insulator. 
The transition does not occur for plumbene in the field range shown in the figure. Higher order effects, due to the stronger spin-orbit coupling in plumbene, emerge at around $E=0.8$ V/\AA~which stops the gap from closing and making the transition. 
The transition for silicene happens at a relatively smaller electric field of $E=0.018$ V/\AA~compared to that of germanene, $E=0.21$ V/\AA, and that of stanene $E=0.56$ V/\AA. 
This suggests that germanene and stanene might be better candidates for any device application based on topological phase transition as silicene might be easily pushed into its trivial regime due to the substrate-induced electric field~\cite{yang_can_2018}. 
We note that the spin splitting stops increasing right at the transition field and saturates to a value equal to the intrinsic band gap. 
The upper bound on the spin splitting results from the intrinsic spin-orbit coupling of the crystal as is shown elsewhere \cite{sunko_maximal_2017}.

\begin{figure}
    \includegraphics[]{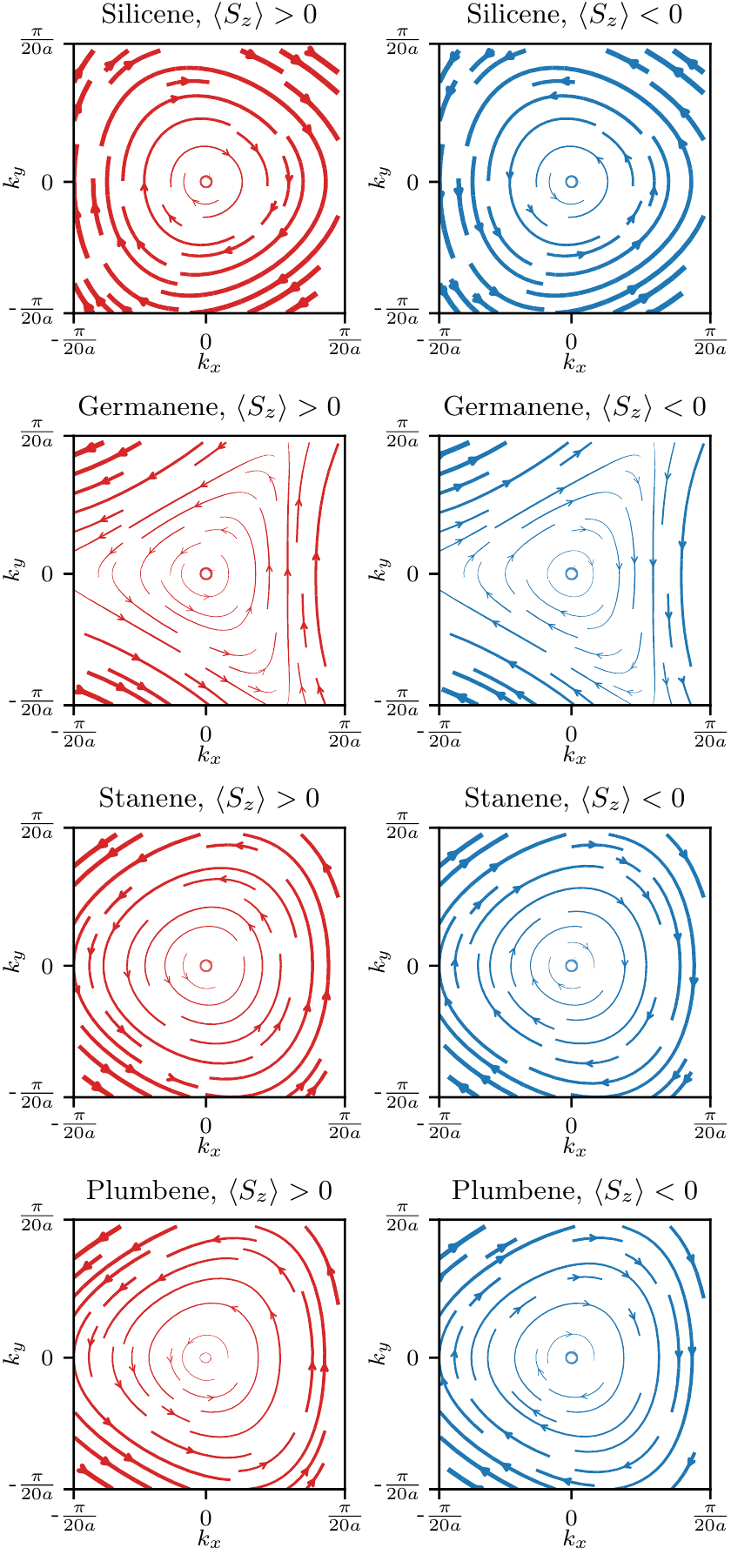}
    \caption[example] 
    {The in-plane spin texture in the vicinity of the $K$ point. The thickness of the curves is proportional to the magnitude of the in-plane spin projection. The electric field is set to $E=0.1$ V/\AA.}
    \label{fig:texture} 
\end{figure}  

The electric field introduces Rashba-like spin-orbit coupling which in turn leads to an in-plane spin texture. Figure \ref{fig:texture} illustrates the expectation value of the in-plane spin projection in the presence of an electric field with the magnitude $E=0.1$ V/\AA~in the vicinity of the $K$. The left and right columns correspond to the two spin-split conduction bands which are highly spin-polarized in the $z$ direction, i.e., $\ev{S_z}\approx \pm\hbar/2$. 
At this specific value of the electric field, silicene, which is in the trivial insulator phase, shows an opposite spin projection to the rest of the monolayers.
Germanene shows a significant trigonal warping in the spin texture which indicates that the higher order spin-orbit terms are more pronounced in germanene than in other monolayers. 
As we will see in Sec. \ref{sec:shc}, these spin orbit terms are responsible for the deviation of the spin Hall conductivity from the quantized value.

\section{\label{sec:invariant}Effective Hamiltonian}
Theory of invariants provide a systematic approach for obtaining an effective Hamiltonian which is expanded to the desired order in terms of the wavevector $\vb*{k}$, the spin $\vb*{s}$, and the external electric field $\vb*{\mathcal{E}}$ \cite{winkler_invariant_2010}. 
In general, the terms appearing in the invariant expansion are the various tensor products of $\vb*{k}$, $\vb*{s}$, and $\vb*{\mathcal{E}}$ that are invariant under the symmetry operations of the point group and, therefore, transform according to the identity representation. 
The coefficients of the invariant expansion are quantified by first-principles calculations or experiments.  
Here we utilize a $4\times4$ effective Hamiltonian which contains invariant terms that are at most linear in $\vb*{k}$, $\vb*{s}$, and $\vb*{\mathcal{E}}$. 
This effective Hamiltonian was derived by Geissler et al.\cite{geissler_group_2013} for silicene in the presence of the spin-orbit coupling and the electric field.
The basis of the Hamiltonian consists of two spin and two sublattice states. 
The effective Hamiltonian resulting from the invariant expansion is $\mathcal{H}(\vb*{k}) = \mathcal{H}_0(\vb*{k}) + \mathcal{H}_\mathcal{E}(\vb*{k})$ where $\mathcal{H}_0(\vb*{k})$ is the Hamiltonian without the external electric field
\begin{equation}
\label{eq:hamil0}
\begin{split}
    \mathcal{H}_0(\vb*{k}) & = a_1\tau_z\sigma_zs_z + a_2(\tau_zk_x\sigma_x + k_y\sigma_y) \\
    & + a_3\sigma_z(s_xk_y - s_yk_x).
\end{split}
\end{equation}
Here $\tau_z=\pm 1$ denotes the $K$ and $K'$ nonequivalent valleys, respectively. The Pauli matrices $\sigma_i$ and $s_i$ operate in the sublattice and spin spaces. 
The coefficients $a_1=\Delta_\text{SO}/2$, $a_2=\hbar v_\text{F}$, and $a_3$ are interpreted as the spin-orbit band gap, the Fermi velocity, and a Rashba-like  spin-orbit coupling term \cite{liu_low-energy_2011, kochan_model_2017} which also introduces some corrections to the Fermi velocity. 
In the current section we do not list the values for $a_3$ but provide the values of the spin-orbit gap and the Fermi velocity. 
The Hamiltonian is modified in the presence of the electric field by $\mathcal{H}_\mathcal{E}(\vb*{k})$ as follows
\begin{equation}
\label{eq:hamile}
\begin{split}
    \mathcal{H}_\mathcal{E}(\vb*{k}) & = a_4\sigma_zs_0\mathcal{E}_z + a_5\tau_z\sigma_0s_z\mathcal{E}_z + a_6\sigma_0(s_xk_y - s_yk_x)\mathcal{E}_z \\ 
    & + a_7(\tau_z\sigma_xs_y - \sigma_ys_x)\mathcal{E}_z 
    + a_8(\sigma_xk_x + \tau_z\sigma_yk_y)s_z\mathcal{E}_z\\
    & + a_9\Big(\sigma_x(s_xk_y + s_yk_x) + \tau_z\sigma_y(s_xk_x - s_yk_y)\Big)\mathcal{E}_z.
\end{split}
\end{equation}
The terms proportional to $a_4$ and $a_5$ represent the electric-field-induced spin splitting. The $a_4$ term introduces a spin splitting only close to the $K$ point whereas the spin splitting by the $a_5$ term is almost independent of the value of $k$. From the first-principles band structure in the previous section we know that the splitting decreases going away from the $K$ point. Therefore, it is reasonable to assume that $a_4\gg a_5$. Hence, we only quantify the $a_4$ coefficient.  
The term $a_6$ represents a Rashba-like energy shift which displaces the spin-polarized bands horizontally. Similarly, since the horizontal shifts in the first-principles band structure are negligible, we have $a_4\gg a_6$.
Finally, the terms $a_7$, $a_8$, and $a_9$ represent higher order corrections to the spin-splitting and the Fermi velocity.
Although symmetry allows many terms in the invariant expansion, 
Four coefficients $a_1$, $a_2$, $a_3$, and $a_4$ are sufficient to describe the four low-energy bands in the vicinity of the $K$ point. 
These coefficients for silicene, germanene, stanene, and plumbene are listed in Table \ref{tb:coeff}. 
\def\arraystretch{1.5}
\begin{table}[b]
\centering
\caption{The coefficients $a_1$--$a_4$ of the effective Hamiltonian given by the invariant expansion in Eqs. \ref{eq:hamil0} and \ref{eq:hamile} and extracted by fitting the DFT results for different monolayers. The units for each coefficient are mentioned in parentheses.}
\label{tb:coeff}
\begin{tabular}{lcccc}
\hline
& Silicene & Germanene & Stanene & Plumbene \\
\hline
$a_1$ (eV) & 0.0007 & 0.0119 & 0.0386 & 0.2385 \\
$a_2$ (eV$\cdot$\AA) & 3.499 & 3.177 & 2.783 & 1.599 \\
$v_\text{F}$ ($10^5$ m/s) & 5.316 & 4.827 & 4.228 & 2.429 \\
$a_3$ (eV$\cdot$\AA) & 0.009 & 0.069 & 0.46 & 0.85  \\
$a_4$ (e$\cdot$\AA) & 0.0417 & 0.0564 & 0.0689 & 0.0589 \\
\hline
\end{tabular} 
\end{table}
These values are mostly in accordance with the ones previously reported \cite{liu_low-energy_2011, tsai_gated_2013}. 
As seen from the table, heavier elements show a relatively larger intrinsic band gap, a lower Fermi velocity, and a stronger spin-orbit coupling.  
However, the electric field induced spin-splitting denoted by $a_4$ shows similar values for different monolayers. 
We note that the values of $a_4$ are considerably smaller than a value one would estimate by assuming a bare electric potential, that is $\Delta_\text{SO}/\mathcal{E}d$. 
This is mainly due to the screening of the electric field by the carriers which in turn leads to carrier redistribution between the sublattice layers and consequently reduces the effectiveness of the external electric field.
The effective Hamiltonian is valid as long as the higher order effects of the electric field are not present, that is $E<0.8$ V/\AA.

We note that there are different numbers reported in the literature for the Rashba parameter of Stanene. For instance Ref. \cite{liu_low-energy_2011} reports a Rashba parameter of $a\lambda_\mathrm{R} = 0.045$ eV$\cdot$\AA~whereas Ref. \cite{tsai_gated_2013} reports $a\lambda_\mathrm{R} = 0.088$ eV$\cdot$\AA. These numbers are an order of magnitude smaller than the value reported in our work, i.e., $0.46$ eV$\cdot$\AA. Although the values obtained for silicene and germanene are comparable to the ones in the literature, the discrepancy in the case of stanene could be resulted from a different methods used to fit the effective model to the ab initio results. Since the $a_3$ term does not change the band gap or the Fermi velocity of the bands, we resorted to the expectation value of in-plane spin operators to fit the $a_3$ parameters which might not be the case in other works. While other parameters of the model are more or less similar to the ones reported before, the Rashba values seem not to be conclusive and require further investigation. 

\section{\label{sec:shc}Spin Hall Conductivity}
The spin Hall conductivity is a linear response coefficient that describes a spin current as the response of the system to an applied electric field.
It can be calculated by using the Kubo formula in terms of a Berry-like curvature $\Omega_{\alpha\beta,n}^\gamma(\vb*{k})$, also called the spin Berry curvature, as follows \cite{gradhand_first-principle_2012}
\begin{equation}
\label{eq:shc}
    \sigma_{\alpha\beta}^\gamma = -\qty(\frac{e^2}{\hbar})\qty(\frac{\hbar}{2e})\int\frac{d^3\vb*{k}}{(2\pi)^3}\sum_{n} f(\epsilon_{n,\vb*{k}})\Omega_{\alpha\beta,n}^\gamma(\vb*{k}),
\end{equation}
where $f(\epsilon_{n,\vb*{k}})$ is the Fermi-Dirac distribution function. 
The spin Berry curvature is given as 
\begin{equation}
\label{eq:berry}
    \Omega_{\alpha\beta,n}^\gamma(\vb*{k}) = \hbar^2\sum_{m\not=n} \frac{-2\Im{\bra{n\vb*{k}}\mathcal{J}_{\alpha}^\gamma\dyad{m\vb*{k}} v_{\beta}\ket{n\vb*{k}}}}{(\epsilon_{n,\vb*{k}} - \epsilon_{m,\vb*{k}})^2}, 
\end{equation}
where $v_\beta$ is the velocity operator and  $\mathcal{J}_{\alpha}^\gamma = \acomm{v_\alpha}{\sigma_\gamma}/2=(v_\alpha\sigma_\gamma + \sigma_\gamma v_\alpha)/2$ is the spin velocity operator. 

Using the effective Hamiltonian derived in the last section, we first calculate the contribution of the Dirac bands at the Fermi level to the spin Hall conductivity. In doing so, we assume that the dominant contribution to the spin Hall conductivity comes from the K point. As we will see, this assumption might not hold true as the low energy levels appear at the $\Gamma$ point as well for stanene. 
Nevertheless, since the electric field induces the band inversion only at the K point, it is still useful to provide an estimate of the spin Hall conductivity from the Dirac bands. 
The Hamiltonian of the system, considering the distant bands in the vicinity of the $K$ point is 
\begin{equation}
    H = 
    \begin{pmatrix}
        H_0(\vb*{k}) & H_{01}(\vb*{k}) \\
        H_{10}(\vb*{k}) & H_1(\vb*{k}),
    \end{pmatrix}
\end{equation}
where $H_0(\vb*{k})$ is the effective Hamiltonian given in Eq. \ref{eq:hamil0}, $H_1(\vb*{k})$ represents the distant bands, and $H_{01}(\vb*{k})$ is the interaction between the low-energy bands and the distant bands. 
Assuming that the distant bands are only interacting weakly, which might not be true in general, it can be shown from the Lowdin partitioning that the effect of the distant bands is quadratic in the leading order. That is, 
\begin{equation}
    \tilde{H}(\vb*{k}) = H_0(\vb*{k}) + \mathcal{O}(H_{01}^2(\vb*{k})). 
\end{equation}
Considering only the $H_0(\vb*{k})$ term, the spin Hall conductivity can be obtained from the Kubo formula as is worked out in \cite{dyrdal_intrinsic_2012}. 
\begin{equation}
\label{eq:naive}
    \sigma_{xy}^z = \frac{a_2^2}{a_2^2 + a_3^2}\frac{e}{4\pi}, 
\end{equation}
where $a_2$ and $a_3$ correspond to the Fermi velocity and the intrinsic spin-orbit coupling and are given in Table \ref{tb:coeff}. 
A spin-orbit coupling term that is much smaller than the Fermi velocity term, i.e., $a_3 \ll a_2$, results in a quantized spin Hall conductivity $\sigma_{xy}^z$ in units of $e/4\pi$ as predicted before in the literature \cite{dyrdal_intrinsic_2012, matthes_intrinsic_2016, matusalem_quantization_2019}.
As we discuss next, the spin Hall conductivity calculated from first principles within the density functional theory is not quantized in $e/4\pi$. 
This discrepancy suggests that the assumptions made in the derivation of Eq. \ref{eq:naive} may not be correct, that is the coupling to the distant bands contributes significantly to the spin Hall conductivity. 
We note that even if the Hamiltonian is completely diagonalized, the contributions from the matrix elements of the velocity operator are still present as the energy eigenstates are not simultaneous eigenstates of the velocity operator as well. 
Therefore, a complete picture of the spin Hall effect in these materials requires us to incorporate a full band Hamiltonian. We do so by performing first principles calculations based on the density functional theory. 
We note that earlier works in the literature have calculated the spin Hall conductivity of germanene \cite{matthes_intrinsic_2016} and stanene \cite{matusalem_quantization_2019} at the Fermi level. 
We go a step further by including  silicene and plumbene in our calculations as well as providing results for a wide range of energies and for several values of the external electric field. 

The spin Hall conductivity is in general a tensor of rank three with 27 components. 
However, one need not calculate all the components as symmetry simplifies the calculations by relating the components to each other. 
The space group of the buckled monolayers, $P\overline{3}m1$, corresponds to the magnetic Laue group $\overline{3}m11'$. According to this symmetry classification, it can be shown that the number of independent tensor components reduces to four \cite{seemann_symmetry-imposed_2015}. 
Out of these four components, $\sigma_{xy}^z=-\sigma_{yx}^z$ dominates the rest. 
This is in fact the component corresponding to the quantum spin Hall effect. 
Here we calculate $\sigma_{xy}^z$ for the buckled monolayers over a wide range of energies. 
To evaluate Eq. \ref{eq:shc}, we use the Wannier interpolation method \cite{wang_abinitio_2006, qiao_calculation_2018,  ryoo_computation_2019} recently implemented in Wannier90 code \cite{qiao_calculation_2018, pizzi_wannier90_2020}.
This method takes advantage of the smoothness of the maximally localized Wannier gauge to integrate the spin-Berry curvature over the Brillouin zone. 
The results are shown in Fig. \ref{fig:shc} where the spin Hall conductivity is in units of $e/4\pi=(e^2/h)(\hbar/2e)$. 
The plots on the left column show $\sigma_{xy}^z$ in the absence of the external electric field and, therefore, the bands are not spin split.
As seen from the figure, the values of $\sigma_{xy}^z$ at the Fermi energy are  not quantized. 
The quantization value is in units of $e/4\pi$.
The values for silicene, germanene, stanene, and plumbene at the Fermi level are $\sigma_{xy}^z=$0.20, 1.05, 1.80, and 3.41, respectively. 
These values suggest that the heavier the element is, the higher spin Hall conductivity it shows at the Fermi energy. 
This behavior is consistent with the previous calculations in 3D topological insulators \cite{matthes_intrinsic_2016, farzaneh_intrinsic_2020}. 
The reason that $\sigma_{xy}^z$ is not quantized is in general due to terms that do not conserve spin $s_z$ as pointed out in the literature \cite{kane_quantum_2005, matusalem_quantization_2019}. 
From the invariant Hamiltonian in Eqs. \ref{eq:hamil0} and \ref{eq:hamile} one can see that $[\mathcal{H}, s_z]\not=0$ due to the existence of Rashba-like terms such as the term with coefficient $a_3$. 
It is worth mentioning that the $a_3$ term is a result of the buckling and is not present in planar structures such as graphene \cite{winkler_invariant_2010} which shows a quantized spin Hall conductivity. 

\begin{figure}[H]
    \includegraphics[]{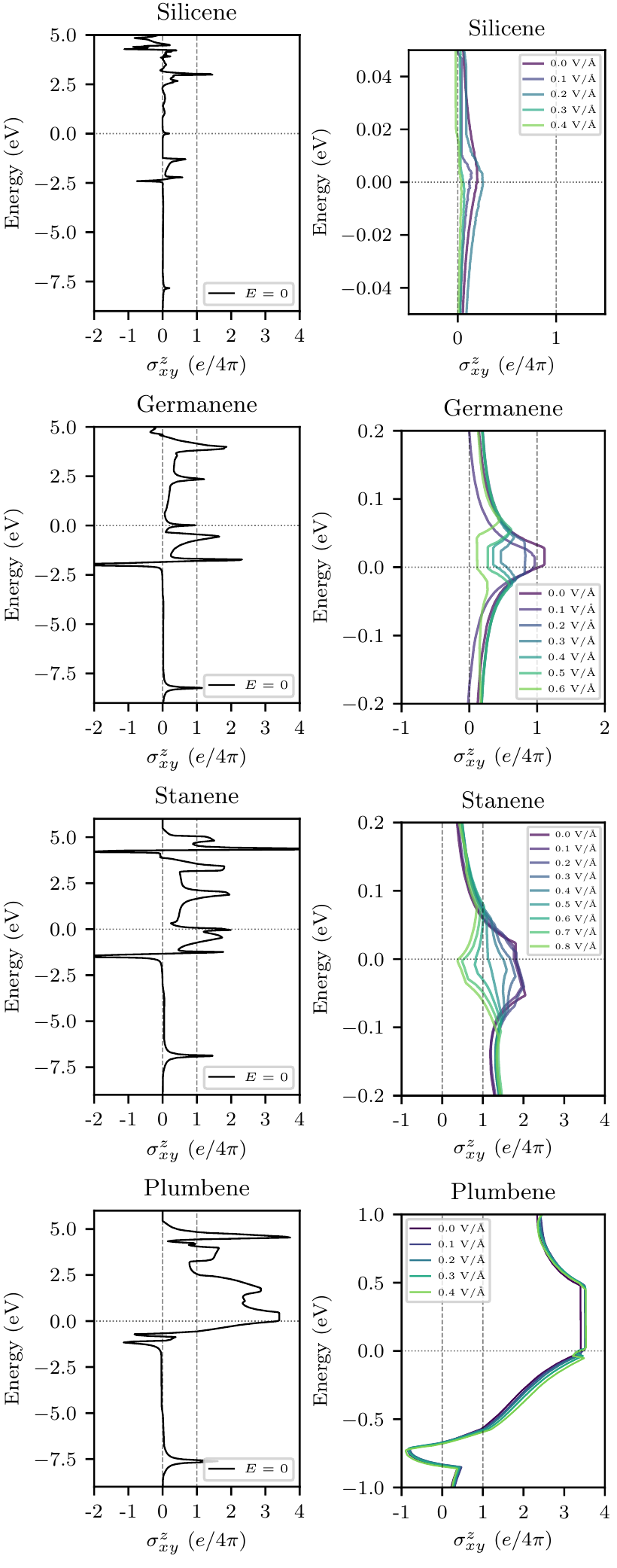}
    \caption[example] 
    {Spin Hall conductivity as a function of the Fermi energy for different values of the electric field for silicene, germanene, stanene, and plumbene. The left column shows the SHC for a wide range of energies whereas the right column contains the zoomed-in plots in the vicinity of the Fermi energy. The electric field is changed in 0.1 V/\AA increments.}
    \label{fig:shc} 
\end{figure}
Although a significant $a_3$ term in Eq. \ref{eq:naive} could explain a spin Hall conductivity that is smaller than $e/4\pi$ in a Dirac band model, it cannot describe the corresponding values of stanene and plumbene which are greater than $e/4\pi$. 
As mentioned this pertains to the fact that the contribution of the other bands below the Fermi level are not negligible. 
Moreover, as seen from Fig. \ref{fig:shc-k}, which illustrates the k-resolved spin Berry curvature, contributions from other points in the Brillouin zone such as the $\Gamma$ point in stanene contribute as well to the spin Hall conductivity. 

The behavior of the spin Hall conductivity in the presence of an electric field depends on the topological phase of the system. 
The right column of Fig \ref{fig:shc} illustrates $\sigma_{xy}^z$ in a close vicinity of the Fermi level for several values of the electric field where the magnitude of the electric field is changed in steps of 0.1 V/$\AA$.
Our general observation is that for electric fields less than the critical value the spin Hall conductivity at the Fermi energy remains the same. As the electric field nears the critical value and goes beyond it, the spin Hall conductivity degrades significantly. 
This can be seen as a result of field-dependent terms in the invariant Hamiltonian in Eq. \ref{eq:hamile} which introduce additional spin mixing and, therefore, degrade the $s_z$ conservation. 
From a lattice point of view, the switching of the topological phase to the trivial phase by the electric field can be interpreted as the point where the asymmetry between the sublattices outweighs the intrinsic spin-orbit coupling. As a consequence, the ground state changes from the linear combination of different sublattices to a single sublattice with both spins. Therefore, the system does not consist of two copies of the integer quantum Hall state \cite{kane_quantum_2005} anymore and the quantum spin Hall phase is destroyed. 
As seen from Fig. \ref{fig:shc} the value of $\sigma_{xy}^z$ at the Fermi energy for silicene, germanene, and stanene, in the presence of an electric field greater than the critical value reduces significantly, whereas the value for plumbene that has no critical value remains the same.
This shows that the topological quantum spin Hall phase and the spin Hall conductivity in plumbene are robust to the perpendicular electric field which can be beneficial for spin generation device applications with robustness to the external field effects. On the other hand, silicene, germanene, and stanene can be suitable candidates for device applications requiring switching between the topological and trivial phases.
It is worth mentioning that the spin Hall conductivity of stanene does not seem to be constant inside the gap as the electric field is increased. The main reason for it is that the the actual gap is smaller than the gap at the $K$ point. This is due to the fact that the valence band at the $\Gamma$ point has a slightly higher energy than the top of the valence band at the $K$ point resulting in an indirect band gap. 
\begin{figure}[H]
    \includegraphics[]{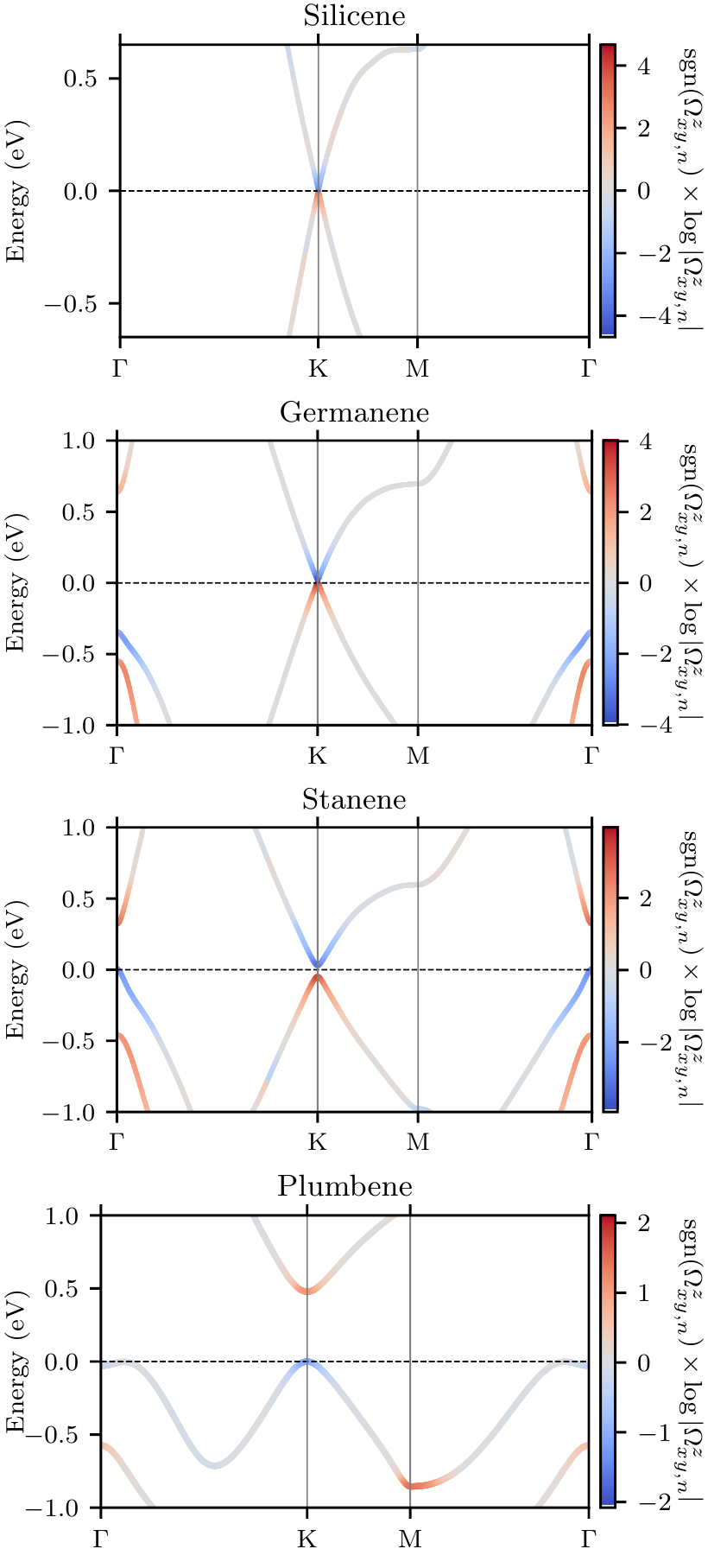}
    \caption[example] 
    {$k$-resolved contribution to the spin-Berry curvature for silicene, germanene, stanene, and plumbene.}
    \label{fig:shc-k} 
\end{figure}  

\section{\label{sec:conclusions}Conclusions}
By performing systematic first-principles calculations based on density functional theory we calculated the fully relativistic band structure of the buckled monolayers of the group 14 with and without an external electric field. 
Various spin properties of the monolayers, such as the spin splitting, spin-dependent band gaps, coefficients of the invariant Hamiltonian, and the spin Hall conductivity, were calculated. 
The main results of this work are the energy-dependent spin Hall conductivity of the buckled monolayers for various values of the electric field.
Our results show that in contrast to the prediction of the simplified two-band Dirac Hamiltonian \cite{dyrdal_intrinsic_2012}, the spin Hall conductivity is not quantized in these monolayers.
Previous ab initio works \cite{matthes_intrinsic_2016, matusalem_quantization_2019} on germanene and stanene report a quantized spin Hall conductivity. Our results show that it could be accidental that the spin Hall conductivity of germanene is close to the quantization value, i.e., 1.05 ($e/4\pi$). The reason is that we do not observe the same quantization for stanene as well as silicene and plumbene. 
We argued that the reason for this discrepancy is the coupling to the remote bands as well as contributions from other points of the Brillouin zone such as the Gamma point. 
Another reason could be the use of a different method by Ref. \cite{matusalem_quantization_2019} for the calculation of the spin Hall conductivity. 
The results also suggest that silicene, germanene, and stanene are suitable for applications involving topological phase transition whereas plumbene, due to the absence of a critical electric field, is more suitable for applications requiring a spin Hall effect that is robust to the external fields. 
It should be noted that there are inherent limitations to the density functional theory such as the underestimation of the band gap which can in principle affect our numerical results. This limitation can be alleviated to an extent by including the self energy using the GW method in the ab initio framework. 
However, the qualitative conclusions of the work are expected to remain the same.
Moreover, verifying numerical results, such as the value of the spin Hall conductivity, requires comparing theory with experimental data. However, since spin Hall effect is difficult to observe directly, experiments involving indirect methods of measuring spin currents, such as the spin-torque ferromagnetic resonance method \cite{liu_spin-torque_2011}, will prove to be valuable in this regard. 

\section*{Acknowledgements}
This material is based upon work funded by AFRL/AFOSR, under AFRL Contract No. FA8750-21-1-0002. The authors also acknowledge the support of NSF through the grant no. CCF-2021230.

\bibliography{ref}

\pagebreak
\clearpage
\onecolumngrid

\setcounter{page}{1}
\setcounter{section}{0}
\setcounter{equation}{0}
\setcounter{figure}{0}
\setcounter{table}{0}
\renewcommand{\thepage}{S\arabic{page}}
\renewcommand{\thesection}{S\arabic{section}}
\renewcommand{\theequation}{\thesection.\arabic{equation}}
\renewcommand{\thefigure}{S\arabic{figure}}
\renewcommand{\thetable}{S\arabic{table}}

\section*{Supplemental Material}

\input{supp}

\end{document}

%% file: supp.tex
In this document we provide an in-depth symmetry analysis of the monolayers of the group 14 of the periodic table namely graphene, silicene, germanene, stanene, and plumbene. 
The character tables of the point groups related to the honeycomb lattice in various forms such as the planar/buckled, with spin-orbit coupling, and in the presence of an electric field are provided as well. 

\subsection{Planar Honeycomb: Graphene}
The crystal structure of graphene consists of two triangular lattices, shifted with respect to each other, which form a honeycomb structure. 
It can also be thought of as a Bravais lattice with a basis of two atoms. 
The space group of graphene can be determined by inspection. 
Since graphene does not have any glide plane or screw axis symmetries, its space group is symmorphic. 
Therefore, with a suitable choice of origin, in this case the center of the hexagons, one can easily determine the symmetry operations and the corresponding point group which is $D_{6h}$ and corresponds to the space group $P6/mmm$ (\#191). 
The point group can be written as the direct product of the point group $D_6$ and the inversion operation $i$, i.e., $D_{6h} = D_6\otimes i$. 
Since we already know, from first-principles calculations and experiments, that the low energy excitations of graphene and the other monolayers reside at the the $K$ point in the Brillouin zone, we need to work with the group of the wave vector at the $K$ point. 
The symmetry of this \textit{little} group is a subgroup of that of the $\Gamma$ point because the 6-fold rotational symmetry is lost as a result of the distinguishability of the $K$ and $K'$ points in a honeycomb structure. 
The group of the wave vector is therefore denoted by the point group $D_{3h} = D_3\otimes \sigma_h$ whose characters are listed in Table \ref{tb:d3h}. Here, one-dimensional representations are denoted by $A$ and the two-dimensional ones by $E$.  
\def\arraystretch{1.2}
\begin{table}[ht]
\centering
\caption{Character table for point group $D_{3h}=D_3\otimes\sigma_h$ which is homomorphic to the group of the wave vector at the $K$ point for space group $P6/mmm$ (\#191).}
\label{tb:d3h}
\vspace{0.1in}
\label{tb:character}
$ \begin{array}{c|rrrrrr}
\hline
& E & 2C_3 & 3C_2' & \sigma_h & 3\sigma_v & 2S_3 \\
\hline
A'_1  & 1 &  1 &  1 &  1 &  1 &  1 \\
A''_1 & 1 &  1 &  1 & -1 & -1 & -1 \\
A'_2  & 1 &  1 & -1 &  1 & -1 &  1 \\
A''_2 & 1 &  1 & -1 & -1 &  1 & -1 \\
E'    & 2 & -1 &  0 &  2 &  0 & -1 \\
E''   & 2 & -1 &  0 & -2 &  0 &  1 \\
\hline
\end{array} $
\end{table} 
The Dirac point in graphene is labeled with the irreducible representation $E''$ which is two-dimensional and therefore implies a 2-fold degeneracy. 

Including spin to the symmetry analysis requires working with double groups which are obtained by a direct product of the spin-less group and the spinor $D_{1/2}$. 
The double group characters for the point group $D_{3h}$, that is $D_{3h}\otimes D_{1/2}$, is listed in Table \ref{tb:d3h-double}.
The new symmetry operation $\mathcal{R}$ is a  $2\pi$ rotation which does not bring a spinor back to its initial state. 
\def\arraystretch{1.2}
\begin{table}[ht]
\centering
\caption{Double group character table for point group $D_{3h}$, i.e. $D_{3h}\otimes D_{1/2}$ where $D_{1/2}$ represents the spinor and transforms according to the irreducible representation $\overline{E}_1$. Here $\mathcal{R}$ denotes a rotation by $2\pi$. The first six irreducible representations are single group and the last three are double group representations.}
\label{tb:d3h-double}
\vspace{0.1in}
\label{tb:character}
$ \begin{array}{c|rrrrrrrrr}
\hline
& E & \mathcal{R} & 2C_3 & 2\mathcal{R}C_3 & 3C_2' & \sigma_h & 3\sigma_v & 2S_3 & 2\mathcal{R}S_3 \\
& & & & & 3\mathcal{R}C_2' & \mathcal{R}\sigma_h & 3\mathcal{R}\sigma_v & \\
\hline
A'_1  & 1 &  1 &  1 &  1 &  1 &  1 &  1 &         1 &         1 \\
A''_1 & 1 &  1 &  1 &  1 &  1 & -1 & -1 &        -1 &        -1 \\
A'_2  & 1 &  1 &  1 &  1 & -1 &  1 & -1 &         1 &         1 \\
A''_2 & 1 &  1 &  1 &  1 & -1 & -1 &  1 &        -1 &        -1 \\
E'    & 2 &  2 & -1 & -1 &  0 &  2 &  0 &        -1 &        -1 \\
E''   & 2 &  2 & -1 & -1 &  0 & -2 &  0 &         1 &         1 \\
\overline{E}_1 & 2 & -2 &  1 & -1 &  0 &  0 &  0 &  \sqrt{3} & -\sqrt{3} \\
\overline{E}_2 & 2 & -2 &  1 & -1 &  0 &  0 &  0 & -\sqrt{3} &  \sqrt{3} \\
\overline{E}_3 & 2 & -2 & -2 &  2 &  0 &  0 &  0 &         0 &         0 \\
\hline
\end{array} $
\end{table}
Since the time-reversal and the inversion symmetries are preserved all the bands at the $K$ point (as well as all other points in the reciprocal space) must be at least doubly degenerate and are labeled with the double group representations, i.e., $\overline{E}_1$, $\overline{E}_2$, and $\overline{E}_3$, which are at least two dimensional. 
On the other hand, since there is no higher dimensional irreducible representations, all the bands at the $K$ points are therefore only 2-fold degenerate. 
This means that the degeneracy of the spinless Dirac point is not preserved as the spin is introduced (no 4-fold degeneracy). And therefore a gap opens up at the Dirac point and one is left with two doubly degenerate bands instead. 
To determine the irreducible representations around the gap, one needs to find the direct product of the $E''$ representation, which denotes the spin-less Dirac point, and the spinor representation $\overline{E}_1$.
The character of a direct product for a symmetry operation $R$ is the product of characters of each representation \cite{dresselhaus_group_2008}. 
\begin{equation}
    \chi^{(\Gamma_i\otimes\Gamma_j)}(R) = \chi^{(\Gamma_i)}(R)\chi^{(\Gamma_j)}(R) 
\end{equation}
The characters of $\chi^{(\Gamma_i\otimes \overline{E}_1)}(R)$ are listed in Table \ref{tb:d3h-direct-pord}.
\def\arraystretch{1.2}
\begin{table}
\centering
\caption{Characters for direct products $\Gamma_i\otimes \overline{E}_1$ for the double group with $D_{3h}$ symmetry where $K_7$ representation corresponds to the spinor $D_{1/2}$.}
\label{tb:d3h-direct-pord}
\vspace{0.1in}
\label{tb:character}
$ \begin{array}{c|rrrrrrrrr}
    \hline
    & E & \mathcal{R} & 2C_3 & 2\mathcal{R}C_3 & 3C_2' & \sigma_h & 3\sigma_v & 2S_3 & 2\mathcal{R}S_3 \\
    & & & & & 3\mathcal{R}C_2' & \mathcal{R}\sigma_h & 3\mathcal{R}\sigma_v & \\
    \hline
    E' \otimes \overline{E}_1 & 4 & -4 & -1 &  1 &  0 &  0 &  0 & -\sqrt{3} &  \sqrt{3} \\
    E'' \otimes \overline{E}_1 & 4 & -4 & -1 &  1 &  0 &  0 &  0 &  \sqrt{3} & -\sqrt{3} \\
    \overline{E}_1 \otimes\overline{E}_1 & 4 &  4 &  1 &  1 &  0 &  0 &  0 &         3 &         3 \\
    \overline{E}_2 \otimes\overline{E}_1 & 4 &  4 &  1 &  1 &  0 &  0 &  0 &        -3 &        -3 \\
    \overline{E}_3 \otimes \overline{E}_1 & 4 &  4 & -2 & -2 &  0 &  0 &  0 &         0 &         0 \\
    \hline
\end{array} $
\end{table}
The direct product representation is in general reducible. One can write them in terms of irreducible representations by using the decomposition theorem which states \cite{dresselhaus_group_2008}
\begin{equation}
    \chi^{(\lambda\otimes\mu)}(R) = \sum_\nu a_{\lambda\mu\nu} \chi^{(\nu)}(R).
\end{equation}
The coefficients are obtained by using the orthogonality theorem as follows  
\begin{equation}
\label{eq:decomposition}
    a_{\lambda\mu\nu} = \frac{1}{h}\sum_{\alpha} N_\alpha \chi^{(\nu)}(C_\alpha)^* \qty[\chi^{(\lambda\otimes\mu)}(C_\alpha)]
\end{equation}
where $N_\alpha$ is the number of elements in class $C_\alpha$ and $h$ is the order of the group, that is the number of its symmetry elements. 
The decomposition of all the products of the form $\chi^{(\Gamma_i\otimes \overline{E}_1)}(R)$ in terms of the irreducible representations is provided in Table \ref{tb:d3h-decompose}. 
\def\arraystretch{1.2}
\begin{table}[ht]
\centering
\caption{Direct products $\Gamma_i\otimes \overline{E}_1$ in terms of irreducible representations of the double group $D_{3h}$ decomposed by using Eq. \ref{eq:decomposition}.}
\label{tb:d3h-decompose}
\vspace{0.1in}
\label{tb:character}
$ \begin{array}{l}
\hline
A'_1 \otimes \overline{E}_1 = \overline{E}_1\\
A''_1 \otimes \overline{E}_1 = \overline{E}_2\\
A'_2 \otimes \overline{E}_1 =\overline{E}_1\\
A''_2 \otimes \overline{E}_1 = \overline{E}_2\\
E' \otimes \overline{E}_1 = \overline{E}_2 + \overline{E}_3\\
E'' \otimes \overline{E}_1 = \overline{E}_1 + \overline{E}_3\\
\hline
\overline{E}_1 \otimes \overline{E}_1 = A_1 + A'_2 + E''\\
\overline{E}_2 \otimes \overline{E}_1 = A''_1 + A''_2 + E'\\
\overline{E}_3 \otimes \overline{E}_1 = E' + E''\\
\hline
\end{array} $
\end{table} 
From this table we see that $E''\otimes \overline{E}_1=\overline{E}_1 + \overline{E}_3$ which means that the Dirac point with the $E''$ representation decomposes into two double group representations, $\overline{E}_1$ and $\overline{E}_3$, after the inclusion of spin-orbit coupling. 

\subsection{Buckled Honeycomb: Silicene, Germanene, Stanene, and Plumbene} 
The buckling in the honeycomb structure breaks some of the symmetries of the original planar structure such as the $C_2$ and $C_6$ rotational symmetries. 
The resulting point group is $D_{3d}$ with the corresponding space group $P\overline{3}m1$ (\#164). 
Since the inversion symmetry is still preserved, this point group can be written as the direct product $D_{3d}=D_3\otimes i$. The characters are listed in Table \ref{tb:d3d}. 
\def\arraystretch{1.2}
\begin{table}[ht]
\centering
\caption{Character table for point group $D_{3d}$ which is homomorphic to the space group $P\overline{3}m1$ (\#162). The upper left quadrant is the character table for point group $D_3$ since $D_{3d}=D_3\otimes i$. The letters $g$ and $u$ denote even and odd transformations, respectively, under the inversion operator.}
\label{tb:d3d}
\vspace{0.1in}
\label{tb:character}
$ \begin{array}{c|rrr|rrr}
\hline
& E & 2C_3 & 3C_2' & i & 2iC_3 & 3iC_2' \\
\hline
A_{1g} & 1 &  1 &  1 &  1 &  1 &  1 \\
A_{2g} & 1 &  1 & -1 &  1 &  1 & -1 \\
E_g & 2 & -1 &  0 &  2 & -1 &  0 \\
\hline
A_{1u} & 1 &  1 &  1 & -1 & -1 & -1 \\
A_{2u} & 1 &  1 & -1 & -1 & -1 &  1 \\
E_u & 2 & -1 &  0 & -2 &  1 &  0 \\
\hline
\end{array} $
\end{table}
Moving to the $K$ point of the Brillouin zone, the group of the wave vector becomes $D_3$ which is a subgroup of $D_{3d}$ and lacks the $i$, $2iC_3$, and $3iC'_2$ symmetry operations. 
The corresponding double group $D_3\otimes D_{1/2}$ is constructed similarly to the previous section by using the orthogonality theorem. The characters of $D_3\otimes D_{1/2}$ are given in Table  \ref{tb:d3-double}. 
\def\arraystretch{1.2}
\begin{table}[ht]
\centering
\caption{Double group character table for point group $D_3$, i.e., $D_3\otimes D_{1/2}$. Here the spinor $D_{1/2}$ transforms according to the irreducible representation $\overline{E}_1$.}
\label{tb:d3-double}
\vspace{0.1in}
\label{tb:character}
\begin{tabular}{cc|cc|rrrrrr}
\hline
\multicolumn{2}{c}{Basis} & & & $E$ & $\mathcal{R}$ & $2C_3$ & $2\mathcal{R}C_3$ & $3C'_2$ & $3\mathcal{R}C'_2$ \\
\hline
$x^2+y^2, z^2$ & & $K_1$ & $A_1$ & 1 &  1 &  1 &  1 &  1 &  1 \\
& $z$, $R_z$ & $K_2$ & $A_2$ & 1 &  1 &  1 &  1 & $-1$ & $-1$ \\
$(x^2 - y^2, xy)$, $(xz, yz)$ & $(x, y)$, $(R_x, R_y)$ & $K_3$ & $E$ & 2 &  2 & $-1$ & $-1$ &  0 &  0 \\
& & \multirow{2}{*}{$K_4\bigg\{$} & $^{1}\overline{E}$ & 1 & $-1$ & $-1$ &  1 &  $i$ & $-i$ \\
& & & $^{2}\overline{E}$ & 1 & $-1$ & $-1$ &  1 & $-i$ &  $i$ \\
& & $K_6$ & $\overline{E}_1$ & 2 & $-2$ &  1 &  $-1$ &  0 &  0 \\
\hline
\end{tabular}
\end{table} 
In the previous section we saw that the bands at the Dirac point are labeled with the $\overline{E}_1$ and $\overline{E}_3$ irreducible representations of the double group $D_{3h}$. 
These representations are in general reducible for the double group $D_3$. 
Therefore, by applying the decomposition formula in Eq. \ref{eq:decomposition}, we can write $\overline{E}_1$ and $\overline{E}_3$ in terms of the irreducible representations of $D_3$ as follows
\begin{equation}
\begin{split}
    & \overline{E}_1 (D_{3h}) \rightarrow \overline{E}_1 (D_3), \\
    & \overline{E}_3 (D_{3h}) \rightarrow \{{}^1\overline{E}+{}^2\overline{E}\} (D_3).\\
\end{split}
\end{equation}
We note that the $K_4 = {}^1\overline{E}+{}^2\overline{E}$ is considered as a two-dimensional representation consisting of two one-dimensional representations that are conjugate to one another. 
This degeneracy is a result of the global inversion and time reversal symmetries which require all the bands throughout the Brillouin zone to be at least doubly degenerate. 
Therefore, the out-of-plane buckling does not lift the degeneracy of $\overline{E}_1 (D_{3h})$ and $\overline{E}_3 (D_{3h})$ bands. 

\subsection{Substrate and electric field} 
Introducing a substrate or applying an electric field along the high-symmetry axis, to the buckled monolayers lowers the symmetry of their crystal further. The global symmetry goes from $D_{3d}$ to $C_{3v}$ whereas the group of the wave vector at the $K$ point goes from $D_3$ to $C_3$ \cite{ribeiro-soares_group_2014, kogan_symmetry_2013}.
Table \ref{tb:c3-double} lists the double group characters for point group $C_3$. 
\def\arraystretch{1.2}
\begin{table}[ht]
\centering
\caption{Double group character table for point group $C_3$ where $\omega=e^{i2\pi/3}$. Here the spinor $D_{1/2}$ is a reducible representation that is decomposed as $D_{1/2}=K_4 + K_5$.}
\label{tb:c3-double}
\vspace{0.1in}
\label{tb:character}
$ \begin{array}{c|rrrrrr}
\hline
& E & \mathcal{R} & C_3 & \mathcal{R}C_3 & C^2_3 & \mathcal{R}C^2_3 \\
\hline
K_1 & 1 &  1 &  1 &  1 &  1 &  1 \\
K_2 & 1 &  1 &  -\omega^* &  -\omega^* & -\omega & -\omega \\
K_3 & 1 &  1 & -\omega & -\omega &  -\omega^* & -\omega^* \\
K_4 & 1 & -1 & \omega^* & -\omega^* &  \omega & -\omega \\
K_5 & 1 & -1 & \omega & -\omega & \omega^* & -\omega^* \\
K_6 & 1 & -1 &  -1 &  1 &  -1 &  1 \\
\hline
D_{1/2} & 2 & -2 & 1 & -1 & 1 & -1 
\end{array} $
\end{table} 
As seen from the table, all the representations are one dimensional which implies that there is no degeneracy at the $K$ point in the presence of an electric field and a spin splitting occurs as a result. 
Decomposition of the irreducible representations of $D_3$ into those of $C_3$ yields 
\begin{equation}
\begin{split}
    \overline{E}_1 (D_3) & \rightarrow K_4 (C_3) + K_5 (C_3),\\
    \{{}^1\overline{E}+{}^2\overline{E}\} (D_3) & \rightarrow K_6 (C_3) + K_6 (C_3).
\end{split}
\end{equation}